\documentclass{jps-cp}
\usepackage{txfonts}
\usepackage{footmisc}
\usepackage{multirow}
\usepackage{graphicx}
\usepackage{color}
\usepackage{rotating}

\title{Classification-Based Analysis of Price Pattern Differences Between Cryptocurrencies and Stocks}

\author{Yu \textsc{Zhang}$^{1}$$^{*}$, Zelin \textsc{Wu}$^{1}$, and Claudio \textsc{Tessone}$^{1}$}

\inst{$^{1}$Blockchain Distributed Ledger Technologies, IfI, University of Zurich, Zurich, Switzerland}

\email{zhangyu@ifi.uzh.ch}

\recdate{February 26, 2025}

\abst{
Cryptocurrencies are digital tokens built on blockchain technology, with thousands actively traded on centralized exchanges (CEXs). Unlike stocks, which are backed by real businesses, cryptocurrencies are recognized as a distinct class of assets by researchers. How do investors treat this new category of asset in trading? Are they similar to stocks as an investment tool for investors? We answer these questions by investigating cryptocurrencies' and stocks' price time series which can reflect investors' attitudes towards the targeted assets. Concretely, we use different machine learning models to classify cryptocurrencies' and stocks' price time series in the same period and get an extremely high accuracy rate, which reflects that cryptocurrency investors behave differently in trading from stock investors. We then extract features from these price time series to explain the price pattern difference, including mean, variance,
maximum, minimum, kurtosis, skewness, and first to third-order autocorrelation, etc., and then use machine learning methods including logistic regression (LR), random forest (RF), support vector machine (SVM), etc. for classification. The classification results show that these extracted features can help to explain the price time series pattern difference between cryptocurrencies and stocks.}

\kword{Cryptocurrency, Stock, Price Time Series, Machine Learning, Classification, Investors, Features}
\begin{document}
\maketitle

\section{Introduction and Related Work}\label{sec1} 

The 2008 financial crisis made people lose much trust in the traditional centralized finance system(CeFi). In 2009, Bitcoin, a blockchain-based payment system, was launched by an author or a group with the pseudonym Satoshi Nakamoto. They published the white paper entitled `Bitcoin: A Peer-to-Peer Electronic Cash System' \cite{nakamoto2008peer} and then built the Bitcoin network in January 2009. Since its beginning, Bitcoin gained much interest from the public because of its innovative characteristics in payment, including decentralization, transparency, and security \cite{sornette2025transaction}. Bitcoin started trading in a centralized exchange, just like any stock in a stock exchange, at the price of \$0.0008 in 2010, which made investors realize that Bitcoins (BTCs) may be valuable. After Bitcoin, plenty of other blockchain platforms were built and related native tokens, called cryptocurrencies, were traded in centralized exchanges, such as Ethereum (ETH), DogeCoin (DOGE), Monero (XMR), Cardano (ADA), et al..

The cryptocurrencies trading market is very huge and continues to grow. For example, the total cryptocurrency trading market size is around 18 trillion dollars in 2024, which marks a 130\% increase in trading volume compared to 2023. Cryptocurrencies are traded within 7$\times$24 hours and there exist many big centralized exchanges that facilitate investors' and traders' cryptocurrency trading, such as Binance, Coinbase, Kraken, Kucoin, et al.. In 2013, only 66 cryptocurrencies were traded in centralized exchanges, while more than ten thousand of cryptocurrencies are being trading as of January, 2025 according to the statistics from \textit{Statistica}\footnote{https://www.statista.com/statistics/863917/number-crypto-coins-tokens/}. 
Although these cryptocurrencies are called currencies, they differ from fiat currencies as they fail to function effectively as a medium of exchange, store of value, or unit of account due to their high price volatility \cite{YERMACK201531}. Though many cryptocurrencies are traded in centralized exchanges, such as Binance, Coinbase, Kraken, et al., unlike stocks, cryptocurrencies' prices are not backed by any firm's businesses. However, they are now regarded as a class of new assets and are traded as an investment vehicle \cite{hong2017bitcoin}.

Based on these analyses, a natural research question about investors lies in whether cryptocurrency investors behave similarly to those investors who trade stocks. Investors' behaviors are reflected in the process of buying and selling of specific asset, which determines the price of the corresponding asset. The price time series records some information about investors' behaviors and mental activities.
So, the research comes to investigate whether cryptocurrencies' price time series hold a different pattern compared to that of stocks during the same period and will fall into the domain of time series classification. If the two kinds of assets' price time series are classified with high accuracy, then we can claim that investors of the two kinds of assets behave differently in trading or in holding the two kinds of assets. Building on this research, the next question is which features explain the differences between the two price time series and to what extent.

Time series classification (TSC) has developed into a well-established field encompassing a wide range of approaches, which include both traditional approaches and machine learning techniques.
Traditional TSC methods include Dynamic Time Warping (DTW) based on distance \cite{muller2007dynamic}; some feature-based methods, like Shapelet Transform, extract features from time series \cite{hills2014classification}; methods such as Hidden Markov Models (HMM) assume inherited processes exist to generate these time series and use probabilistic models \cite{blunsom2004hidden}. 

Machine learning methods significantly enhance data representation and the ability to handle complex time series data. For example, the Convolutional Neural Network (CNN) \cite{lecun1998gradient}, known for its remarkable success in the field of computer vision \cite{krizhevsky2012imagenet} has been effectively applied to time series classification which uses the convolutional layers to capture local temporal dependencies and repeating patterns in the data \cite{wang2017time}. Recurrent neural networks (RNNs) \cite{rumelhart1986learning}, including their variants Long Short-Term Memory (LSTM) \cite{hochreiter1997long} and Gated Recurrent Units (GRU) \cite{dey2017gate}, capture complex dependencies in time series by retaining information across multiple time steps that allow them to recognize long-term patterns and temporal relationships \cite{cho2014learning}. 

In this paper, we will mainly use different machine learning methods to classify cryptocurrencies' and stocks' price time series and then extract features to explain their differences. The organization of the paper is as follows: Data collection and processing are described in Section 2; the classification model is briefly introduced in Section 3; then in Section 4 and 5, we apply machine learning models to classify price time series and extract different features to explain these differences; at last, we conclude and discuss our research.

\section{Data Collection and Processing}

The price time series data for cryptocurrencies and stocks was collected at minute-level, spanning from June 1, 2023, to May 31, 2024. 
735 stocks' price time series from the NASDAQ and NYSE exchanges were collected from Bloomberg\footnote{https://www.bloomberg.com}. 
131 popular cryptocurrency price time series data were collected from Binance with USDT as a unit. The USDT is a stablecoin that pegs the US dollar. One USDT is almost one US dollar. For each asset, four key price features (open, high, low, and close) are collected.

The cryptocurrency market operates 24/7, while the stock market functions only during trading hours on business days. To align their price time series, we select cryptocurrency price points exclusively from 9:30 AM to 4:00 PM (New York time) on working days, matching stock market hours. This ensures price data from different assets is classified within a consistent temporal framework. 
The original stock price time series data have some missing values we filled them with neighboring values from the same day. Each time series consists of 98,532 time points (391 minutes per day × 252 days) at the minute level after selection and filling. 
Then each cryptocurrency and stock price time series with 98,532 time points at the minute level was divided into daily and weekly segments, each referred to as a sample.

For daily segments, each original price time series was divided into 252 samples, with each sample containing 391 time points.
There are 33,012 ($=252\times$131) daily samples for cryptocurrency and 185,220 
($=252\times$735) daily samples for stocks. Z-score normalization was applied for each daily sample, resulting in sample data having a mean of 0 and a variance of 1. 
For weekly segments, each original price time series was divided into weekly segments. Given that some trading weeks contain four days, while others have five, only those weeks that have five trading days were selected. Each weekly segment includes 1,955 ($=391\times5$) data points. After segmentation and selection, there are 5,502 weekly samples for cryptocurrency and 30,970 weekly samples for stock. In the following sections, different machine learning models are applied to the cryptocurrencies' and stocks' daily and weekly samples for classification.

\section{Classification Models}

Nine different machine learning models will be used to classify stock and cryptocurrency price time series data, including \textit{Multilayer Perceptron} (MLP), \textit{Convolutional Neural Networks} (CNN), \textit{Residual Network} (ResNet), \textit{Recurrent neural network} (RNN), \textit{Gated Recurrent Unit} (GRU), \textit{Long-Short-Term Memory} (LSTM), Autoencoder, Time CNN, MultiChannel CNN, whose architecture are shown in Table \ref{tab:architecture} and corresponding critical details are described below.

\begin{table}[!ht]
\centering
\renewcommand{\arraystretch}{1.5}
\footnotesize 
\begin{tabular}{cc}
\hline
\textbf{Method}  & \textbf{Architecture}   \\ \hline
MLP & FC(32)-FC(64)-FC(64)-FC(128)-FC(1) \\ 
CNN & CONV(32)-CONV(64)-CONV(64)-CONV(128)-FC(1) \\ 
ResNet & CONV(64)- Resblock(CONV(64)-CONV(64)*6-FC(1)  \\ 
RNN & RNN(32)-RNN(32)-FC(1)  \\ 
GRU & GRU(32)-GRU(32)-FC(1) \\ 
LSTM & LSTM(32)-LSTM(32)-FC(1)  \\ 
Autoencoder & CONV(64)-CONV(128)-CONV(256)-FC(256)-CONV(256)-CONV(128)-CONV(64)-FC(1)\\ 
Time-CNN & CONV(6)-CONV(12)-FC(1)  \\ 
Multi-Channel-CNN & (concatenate CNN(32), CNN(64), CNN(128)-FC(64)-FC(1) \\ \hline
\end{tabular}
\caption{The architecture of machine learning models applied in classification.
FC: fully connected layer; CONV: 1D convolutional layer.}
\label{tab:architecture}
\end{table}

MLP, established in 1958 \cite{rosenblatt1958perceptron} and refined in the 1980s \cite{rumelhart1986learning}, is acknowledged as a fundamental model in classification. Its hidden layers in MLP are calculated by: 

\begin{equation}
h_{1} = \text{ReLU}(W_{1} x + b_{1}) 
\label{eq:mlp}
\end{equation}

$W_{1}$ and $b_{1}$ represent the weight matrix and bias vector for the first layer, $x$ is the input to the first layer, and $h_1$ is the hidden state (output) of the first layer, $\text{ReLU}$ denotes the rectified linear unit activation function. For calculating the outputs of other layers, the corresponding inputs are the hidden states of their previous layers, and similar formulations are:  $h_{j} = \text{ReLU}\left(W_{j-1} h_{j-1} + b_{j-1}\right)$, where 
$h_{j-1}$ is the hidden states of the $(j-1)^{th}$ layer and $h_j$ is the hidden states of the $j^{th}$ layer. If a price time series is $p=\left(p_1, p_2, \cdots, p_T\right)$ where $T$ indicates the length of the time series, then $p$ can be taken as the input vector of the first layer directly when MLP is used in time series classification, namely $x=p$ in equation (\ref{eq:mlp}).

CNN employs distinct convolutional layers for TSC (1D). The hidden layer of a one-dimensional convolutional network is calculated by:
\begin{equation}
y_i = \sum_{j=1}^{k} w_j \cdot x_{i \cdot s + j - 1} + b
\label{eq:conv}
\end{equation}

After padding, the input sequence $x$ becomes $[p_o, ..., p_1, x'_1, x'_2, ..., x'_l, p'_1, ..., p'_o]$, where $o$ represents the padding size. Padding values (typically zero) are added to the beginning and end of the original input sequence, and $x'_n$ denotes the $n$-th value of the original input. $l$ is the length of the original input, $w=[w_1, w_2, ..., w_k]$ denotes the kernel, $k$ is the kernel size. $y=[y_1, y_2, ..., y_m]$ is the output sequence, $m$ is the length of the output sequence. $s$ denotes stride, $i$ is the index of the output sequence, and $j$ is the index of the kernel. The length of the output sequence of the 1D convolutional layer, $m$ is determined by $m = \left \lfloor \frac{n + 2p - k}{s}\right\rfloor + 1$.

The key feature of ResNet is residual learning (or residual connections) \cite{he2016deep}, which skips connections to jump over one or more layers to facilitate the direct propagation of gradients throughout the network, particularly in deep architectures that experience vanishing gradients. The residual block works according to the following formula:

\begin{equation}
h_{1} = \text{ReLU}(\text{conv}(\mathbf{x}))
\label{eq:res1}
\end{equation}
\begin{equation}
h_{2} = \text{ReLU}(\text{conv}(h_{1}))
\label{eq:res2}
\end{equation}
\begin{equation} h^{*} = \text{ReLU}(h_{2} + \mathbf{x})
\label{eq:res3}
\end{equation}

Where $\mathbf{x}$ is the input to the block, $h_{1}$ and $h_{2}$ are the output of the convolutional layers within the block, $h^{*}$ is the output of the block. 

In RNN, the hidden state at a specific time step combines information from the corresponding input and the previous hidden state, which enables this architecture to record sequence temporal dependencies. The output of a RNN layer is calculated by:
\begin{equation}
h_t = \tanh(W_x x_t + W_h h_{t-1} + b)
\label{eq:rnn}
\end{equation}

where $h_t$ is the hidden state at time step $t$. $x_t$ is the input vector at the time step $t$, $W_x$ is the weight matrix for the input to the hidden state, $W_h$ is the weight matrix for the previous hidden state to the hidden state, and $b$ is the bias, $\tanh$ stands for hyperbolic tangent activation.

GRU mitigated the constraints of typical RNN in training, including the vanishing gradient problem \cite{bengio1994learning} and memory limitation. It improves the effectiveness of the model by including gating techniques that control the information flow of the network.
The update gate regulates how much information from the previous hidden state is preserved in the new hidden state, and the reset gate controls the extent to which the previous hidden state is discarded when generating the candidate hidden state.

LSTM networks enhance traditional RNNs by incorporating gating mechanisms, thus mitigating vanishing gradients and allowing long-term dependency learning. It introduces a cell state that serves as long-term memory and employs three gates: the forget gate discards parts of the previous cell state, the input gate updates the cell state, and the output gate determines the hidden state. 

Autoencoder is an unsupervised learning technique for learning efficient data representations without labels \cite{ng2011sparse}. Autoencoder includes encoder phase and decoder phase. The encoder phase compresses the input into a compact representation, and the decoder phase reconstructs the original input from the compact representation.

Time-CNN \cite{ismail2019deep} includes two consecutive convolutional layers without padding, each followed by a local average pooling layer of size 3; This architecture adheres to the original Time-CNN design using the mean squared error (MSE) as the loss function. However, the ReLU activation function was applied for all convolutional layers in this model, which deviates from the original design that uses sigmoid.

The Multi-Channel Convolutional Neural Network (m-CNN) \cite{ismail2019deep} concatenates the output of the convolutional layers with different kernel sizes. In the m-CNN model used in this paper, the input passes through three convolutional layers with distinct filter sizes (5, 7, and 9) and different numbers of neurons (32, 64, and 128). The “same” padding is used for three convolutional layers to preserve the input dimensions and keep temporal information, which is crucial for subsequent concatenation operations. A max-pooling layer with pooling size 2 succeeds in each convolutional layer, decreasing dimensionality while maintaining critical features. The output of these three max-pooling layers is concatenated. The combined output is further flattened and linked to a fully connected layer, followed by a concluding dense layer using a sigmoid activation function to get the final class prediction.

\section{Classification Experiment and Results}

The experiments for classification between cryptocurrencies and stocks in this section consist of eight main sub-experiments that vary in temporal segmentation (daily versus weekly), different features used for training (all four features versus 'close' only), and data balance (trained data samples are balanced versus not balanced). For imbalanced datasets, we evaluated the classification results using the F1 score, and classification accuracy rate as metrics. For balanced samples, only the accuracy rate is applied in the evaluation. 

For each sub-experiment configuration, the data set was randomly divided into train, validation and test sets, with 80\% of the data allocated for training and the remaining 20\% for testing. From the training set, 20\% of the data was further separated for validation. The same proportion of labels was preserved throughout the train and test sets, ensuring that any class imbalances inherent in the data set were reflected throughout the experimental process. After splitting the data, Z-score normalization was applied for the price data, ensuring that the data had a mean of zero and a standard deviation of one so that the machine learning models could perform more stably.

The classification accuracy rate and F1-score in price time series classification between cryptocurrencies and stocks by using these machine learning methods are extremely high based on the results of Sections \ref{4.1} and \ref{4.2}. Do these results stem from differences in the underlying price time series patterns between cryptocurrencies and stocks or from the superior classification performance of the machine learning methods? 
To answer this question, in Subsection \ref{4.3}, another two complementary experiments are conducted, namely, repeating classification with these nine machine learning models on only cryptocurrencies price time series data and on only stock price time series data, respectively. Concretely, for cryptocurrency price time series data, we randomly set half of these cryptocurrencies' labels as zero and the rest cryptocurrencies' labels as one. Then, the nine machine learning methods are trained and used for classification between cryptocurrencies labeled as zero and one. The same procedures are also applied to the stock price time series. If classification accuracy in the two experiments still remains exceptionally high, the results in Subsections \ref{4.1} and \ref{4.2} can not provide sufficient evidence that price time series patterns differ between cryptocurrencies and stocks; otherwise, such a conclusion can be justified.

Different loss function was used in different sub-experiments. Binary Cross Entropy was applied for most configurations; Mean Squared Error was applied for time CNN. Focal Loss\cite{ross2017focal} was applied for some models training with the whole dataset, especially using daily data. Adam was used to optimize these models\cite{diederik2014adam}. We adjusted the hyperparameters by grid search. Most of the time, the default hyperparameter set Table\ref{tab:hpr} performs well enough. Additionally, we trained these models with different numbers of kernels and different numbers of layers to validate their robustness.

\begin{table}[tbh]
\caption{The default hyperparameter set for experiment}
\label{tab:hpr}
\centering
\begin{tabular}{|c|c|}
\hline
\textbf{Hyperparameter} & \textbf{Default value} \\ \hline
(Initial) Learning rate      & 0.001     \\ \hline
Loss function & BCE     \\ \hline
Batch size & 128  \\ \hline
Training epoch & 500    \\ \hline
Dropout rate(if applied)& 0.2      \\ \hline
\end{tabular}
\end{table}

Each sub-experiment is repeated several times and the evaluation metrics are shown on average below.

\subsection{Experiment on Unbalanced Data Samples (All Data) and Results}\label{4.1}

Firstly, we train models based on all unbalanced data samples. Because the experiments are also run with the configuration of two kinds of different data temporal segmentation and two kinds of data features, then there are four sets of sub-experiments.
Table \ref{tab:f1-results} provides a detailed comparison of classification performance with four different configuration combinations when all data are used in training. Both the F1 score and the accuracy rate are used to measure overall classification performance. 

\begin{table}[h]
\centering
\renewcommand{\arraystretch}{1.5}
\footnotesize
\begin{tabular}{cccccccccc}
\hline
\textbf{Model} & \multicolumn{2}{c}{\textbf{Weekly - 4 features}} & \multicolumn{2}{c}{\textbf{Weekly - 1 feature}} & \multicolumn{2}{c}{\textbf{Daily - 4 features}} & \multicolumn{2}{c}{\textbf{Daily - 1 feature}} \\ \hline
 & \textbf{F1} & \textbf{Acc} & \textbf{F1} & \textbf{Acc} & \textbf{F1} & \textbf{Acc} & \textbf{F1} & \textbf{Acc} \\ \hline
MLP & 98.28\% & 99.48\% & 94.42\% & 98.33\% & 95.08\% & 98.50\% & 83.27\% & 95.48\% \\ 
CNN & 99.43\% & 99.83\% & 93.27\% & 98.07\% & 99.24\% & 99.77\% & 84.97\% & 96.09\% \\ 
ResNet & 99.41\% & 99.82\% & 93.86\% & 98.26\% & 98.50\% & 99.57\% & 89.38\% & 97.37\% \\ 
RNN & 98.83\% & 99.64\% & 95.07\% & 98.54\% & 60.89\% & 89.20\% & 89.03\% & 96.62\% \\ 
GRU & 99.14\% & 99.74\% & 97.48\% & 99.24\% & 96.15\% & 99.11\% & 92.82\% & 98.22\% \\ 
LSTM & 99.17\% & 99.75\% & 97.53\% & 99.25\% & 95.78\% & 99.00\% & 91.78\% & 97.96\% \\ 
Autoencoder & 99.43\% & 99.83\% & 99.20\% & 99.76\% & 97.68\% & 99.50\% & 94.35\% & 98.59\% \\ 
t-CNN & 98.89\% & 99.66\% & 97.12\% & 99.13\% & 95.30\% & 98.66\% & 88.66\% &  96.61\% \\ 
m-CNN & 99.08\% & 99.72\% & 98.24\% & 99.47\% & 96.03\% & 98.97\% & 90.30\% & 97.52\% \\ \hline
\end{tabular}
\caption{Overall accuracy rate (Acc) and F1 score (F1) for four different experiment sets (daily versus weekly segmentation, four features versus one feature) with unbalanced data. The column $Weekly-4features$ means the accuracy rates of models trained on the weekly price time series data with four features. Other columns have a similar meaning.}
\label{tab:f1-results}
\end{table}

As shown in Table \ref{tab:f1-results}, the classification accuracy rate and F1 score of all models are extremely high when weekly slices of price sample data are used. The smallest accuracy rate and F1 score are around 98\% and 93\%, respectively for weekly data. Models' classification accuracy rates and F1 scores, when all four price features (close price, open price, the highest price, and the lowest price) are used, are only a bit higher than that when only one feature (close price) is used, which means that the feature 'close price' already contains the most necessary information about investors behaviors. Different models' classification performances are very similar, which may reflect that cryptocurrency price time series have very different patterns from the stock price time series.

However, the variance of F1 score and accuracy rate from models with daily data is relatively larger than those with weekly data, which means that pattern differences are clearer in longer price time series than in shorter price time series.

\subsection{Experiment on Balanced Data Samples and Results}\label{4.2}

The number of stock data points is almost 7 times that of cryptocurrency, which makes the total sample data extremely unbalanced. To avoid the problem of data imbalance in the classification performance, the stock data was sampled randomly to keep the two kinds of data balanced. After sampling, there are also four sets of sub-experiments for the balanced data samples similarly. 

\begin{table}[h]
\centering
\renewcommand{\arraystretch}{1.5}
\footnotesize 
\begin{tabular}{ccccc}
\hline
\textbf{Model}  & \textbf{Weekly - 4 features} & \textbf{Weekly - 1 features} & \textbf{Daily - 4 features} & \textbf{Daily - 1 features}  \\ \hline
MLP & 98.59\% & 97.18\% & 97.75\% & 93.06\% \\ 
CNN & 99.91\% & 98.64\% & 99.67\% & 92.40\%\\ 
ResNet & 99.23\% & 96.05\% & 99.65\% & 95.43\%\\ 
RNN & 99.41\% & 98.27\% & 98.08\% & 94.48\% \\ 
GRU & 99.64\% & 98.23\% & 98.95\% & 95.68\% \\ 
LSTM & 99.59\% & 99.00\% & 98.78\% & 93.87\% \\ Autoencoder & 99.68\% & 99.50\% & 99.20\% & 97.44\% \\ 
t-CNN & 99.72\% & 97.91\% & 97.66\% & 94.89\% \\ 
m-CNN & 99.46\% & 99.41\% & 98.52\% & 95.96\% \\ \hline
\end{tabular}
\caption{Overall accuracy rates under four sub-experiment configurations (daily versus weekly segmentation, four features versus one feature) with balanced data. The column $Weekly-4features$ means the accuracy rates of models trained on the weekly price time series data with four features. Other columns have a similar meaning. }
\label{tab:acc-results}
\end{table}

Table \ref{tab:acc-results} outlines the results of the machine learning models in the four configurations of experiments using accuracy rate as the evaluation metric. All models have extremely high accuracy rates with $>96\%$ in most sub-experiments. Weekly and daily models' accuracy rates with four features are relatively higher than their corresponding models with only one closed price feature, which means that the closed price may already contain the most relevant information. By comparing models' performance under different empirical segmentation, we find that models trained by weekly data have higher accuracy rates than those models that are trained only by daily data, which means that longer price time series can relatively provide more information about price pattern difference and also investors' trading behaviors.

Machine learning methods are prone to over-fitting by using very complex structures in classification, especially when data are similar to each other. To prove that the extremely high performance of different classification models is due to the clear pattern difference between the cryptocurrency price time series and stock price time series, we do the robust check for classification by training corresponding machine learning models with more complex structures and simpler structures under the same data setting as that in Subsection \ref{4.1} and \ref{4.2}. In the robust check, we still only focus on the accuracy rate because the balanced data is used.

The classification results are shown in Table \ref{tab:robust}. The first column describes the base model structure which is described in Table \ref{tab:architecture}. Each integer denotes the number of neurons in the corresponding layer and each integer denotes a layer. ``Alt Model 1" refers to models with a simpler structure, while ``Alt Model 2" denotes more complex models with additional layers and neurons per layer. ``Acc" means the corresponding models' accuracy rate in classification. We find that all models have very similar performance in classification by comparing the baseline models' accuracy rates to the other two kinds of models' accuracy rates. This robust checking denotes that the high classification accuracy rates in classifying cryptocurrencies' price time series and stocks' price time series are mainly because of the distinct price patterns between these two kinds of assets.

\begin{table}[h]
\centering
\renewcommand{\arraystretch}{1.5}
\footnotesize 
\begin{tabular}{ccccc}
\hline
\textbf{Model}  & \textbf{Baseline Model /Acc} & \textbf{Alt Model 1/Acc} & \textbf{Alt Model 2/Acc} \\ \hline
\multirow{2}{*}{MLP} & 32-64-64-128 & 32-32-64 & 32-64-128-256-512 \\
& 97.75\% & 97.64\% & 97.68\% \\ 
\multirow{2}{*}{CNN} & 32-64-64-128 & 32-32-64 & 32-32-64-64-128-128  \\ 
& 99.67\% & 99.27\% & 99.60\% \\ 
\multirow{2}{*}{ResNet} & 64-64-64-64-64-64 & 32-32-32-32 & 32-32-32-64-64-64-128-128-128 \\ 
& 99.65\% & 98.20\% & 99.41\% \\ 
\multirow{2}{*}{LSTM} & 32-32 & 8-16 & 32-64-64-128  \\ 
& 98.79\% & 98.67\% & 99.01\% \\ 
\multirow{2}{*}{GRU} & 32-32 & 8-16 & 32-64-128  \\ 
& 98.95\% & 98.54\% & 98.52\% \\  \hline
\end{tabular}
\caption{Classification results on models' robust checking under the configuration with daily data and four features on balanced data. The ``Baseline configuration" denotes models that have the same structure as in Table \ref{tab:architecture}. "Alt Model 1" refers to models with a simpler structure, while "Alt Model 2" denotes more complex models with additional layers and neurons per layer. ``Acc" means the corresponding models' accuracy rate in classification.}
\label{tab:robust}
\end{table}

\subsection{Complementary Experiments on Cryptocurrency Data and Stock, Respectively, and Results}\label{4.3}

The Subsections \ref{4.1} and \ref{4.2} show that the classification accuracy rate and the F1 score between the cryptocurrency price time series and the stock price time series are extremely high. To determine where the extremely high accuracy rate and F1 score originate from, the machine learning methods' superior classification performance or the difference in price time series patterns between cryptocurrencies and stocks, two complementary classification experiments are conducted below.

The first complementary experiment is conducted with only cryptocurrency price time series data. Firstly, half number of the 131 cryptocurrencies are randomly selected and labeled as 0, and the rest cryptocurrencies are labeled as 1. Secondly, the 9 machine learning models are trained and used for classification predicting on labeled cryptocurrency price time series. The above procedures are run multiple times to obtain the performance on average. The second complementary experiment is conducted similarly.

As shown in Table \ref{tab:background-crptocurrency}, Table \ref{tab:background-stock},
the results of the two complementary experiments were consistently poor, with the test accuracy rate being around 50\% on average in all models. The low accuracy rates in the two complementary experiments prove that the extremely high accuracy rate and the F1 score in the classification between cryptocurrency price time series and stock price time series are due to their price pattern differences, not due to machine learning models.

\begin{table}[h]
\centering
\renewcommand{\arraystretch}{1.5}
\footnotesize 
\begin{tabular}{ccc}
\hline
\textbf{Method} & \textbf{Train Accuracy} & \textbf{Test Accuracy}  \\ \hline
MLP & 70.37\% & 50.69\%  \\ 
CNN & 61.68\% & 50.29\%  \\ 
ResNet & 95.67\% & 51.71\%  \\ 
RNN & 61.47\% & 50.76\%  \\ 
GRU & 53.98\% & 50.92\%  \\ 
LSTM & 80.32\% & 50.21\%  \\ 
Autoencoder & 50.10\% & 50.50\%  \\ 
t-CNN & 58.75\% & 50.71\%  \\ 
m-CNN & 50.32\% & 50.00\%  \\ \hline
\end{tabular}
\caption{Average results of the complementary experiment using cryptocurrency price time series. The average test accuracy in all machine learning models are approximate 50\%.}
\label{tab:background-crptocurrency}
\end{table}

\begin{table}[tbh]
\centering
\renewcommand{\arraystretch}{1.5}
\footnotesize 
\begin{tabular}{ccc}
\hline
\textbf{Method} & \textbf{Train Accuracy} & \textbf{Test Accuracy}  \\ \hline
MLP & 49.98\% & 50.00\%  \\ 
CNN & 55.28\% & 52.98\%  \\ 
ResNet & 77.35\% & 50.64\%  \\ 
RNN & 53.70\% & 50.72\%  \\ 
GRU & 49.65\% & 50.00\%  \\ 
LSTM & 49.80\% & 50.00\%  \\ \
Autoencoder & 50.12\% & 50.00\%  \\ 
t-CNN & 55.08\% & 52.29\%  \\ 
m-CNN & 96.01\% & 50.67\%  \\ \hline
\end{tabular}
\caption{Average results of the complementary experiment using stock price time series. The average test accuracy in all machine learning models is approximately 50\%.}
\label{tab:background-stock}
\end{table}

\section{Extracting Features to Explain Price Pattern Differences}

In the previous sections, we employed machine learning models to classify stock and cryptocurrency price time series, achieving strong performance and indicating distinct patterns between the two. In this section, we extract and analyze key features from the time series to identify the sources of these differences. The features extracted in this study are primarily human-interpretable, meaning their significance can be understood from their definitions.

\subsection{Return Rate}

The return rate is a central financial metric that measures the profitability of an investment over a specific period by calculating the percentage change relative to its initial value. It is a basis for comparing the performance of various investments and is calculated as $R_{t} = \frac{p_{t} - p_{t-1}}{p_{t-1}}$,
where $t$ denotes the time, $p=[p_1,p_2,...,p_T]$ is the price time series, $T$ is the length of the sequence, $p_{t-1}$ is the price in the previous time step and $p_{t}$ is the price at the time step $t$.

Classification of return rate time series using nine machine learning methods still yields exceptionally high performance, indicating that these series retain the key distinguishing characteristics between cryptocurrency and stock price time series, which is shown in Table \ref{tab:daily-close-return-result}.

\begin{table}[h]
\centering
\renewcommand{\arraystretch}{1.5}
\footnotesize 
\begin{tabular}{ccc}
\hline
\textbf{Method}  & \textbf{Accuracy with Return Rate} & \textbf{Accuracy with Price}   \\ \hline
MLP & 91.51\% & 93.13\% \\ 
CNN & 95.43\% & 94.15\% \\ 
ResNet & 95.64\% & 95.27\% \\ 
RNN & 95.77\% & 95.12\% \\ 
GRU & 96.56\% & 96.07\% \\ 
LSTM & 97.05\% & 94.61\% \\ 
Autoencoder & 97.83\% & 97.50\% \\ 
t-CNN & 95.19\% & 94.21\% \\ 
m-CNN & 96.58\% & 96.27\%\\ \hline
\end{tabular}
\caption{Accuracy of machine learning models on balanced return rate data (using daily data, return rate is calculated using the “close price"). The experiment for calculating the accuracy rate with price is the same as that in Subsection \ref{4.2}. And the experiment in calculating the accuracy with the return rate use the similar settings in data splitting and model training.}
\label{tab:daily-close-return-result}
\end{table}

\subsection{Extracting Other Discrete Features for Classification}

In addition to the return rate, we also extracted several discrete features to better understand and explain the differences in the time series data of the two asset classes, including mean, variance, maximum, minimum, kurtosis, skewness, and first to third-order autocorrelation, the count of maximum and minimum peaks etc. which are shown and defined in Table \ref{tab:feature-summary} in detail. All features are calculated based on the daily price time series data.

\begin{sidewaystable}[tbh]
\centering
\renewcommand{\arraystretch}{1.3}
\footnotesize 
\begin{tabular}{l l p{8cm}} 
\hline
\textbf{Feature Category} & \textbf{Feature Name} & \textbf{Description} \\ \hline
\multirow{4}{*}{\textbf{Statistical Features}} 
    & Mean & Measures the central tendency of a time series. \\
    & Variance & Measures the dispersion of a time series. \\
    & Maximum (Max) & Highest value of a time series. \\
    & Minimum (Min) & Lowest value of a time series. \\
    & Kurtosis & Measures the tailness of the distribution. \\
    & Skewness & Measures the asymmetry of the distribution. \\
\hline
\multirow{3}{*}{\textbf{Autocorrelations}} 
    & 1st Order Autocorrelation & Measures correlation between adjacent time points in a time series. \\
    & 2nd Order Autocorrelation & Measures correlation among time points with a two-step lag in a time series. \\
    & 3rd Order Autocorrelation & Measures correlation among time points with a three-step lag in a time series. \\
\hline
\multirow{2}{*}{\textbf{Difference-Based Features}} 
    & Mean of Differences & Average of successive differences in the sequence. \\
    & Mean of Absolute Differences & Average of absolute differences between successive values. \\
\hline
\multirow{3}{*}{\textbf{Signal Characteristics}} 
    & Peak-to-Peak Distance & Difference between the highest and lowest peaks. \\
    & Area Under the Curve (AUC) & Integral of the time series, representing accumulated magnitude. \\
    & Entropy & Measures randomness or unpredictability of the sequence. \\
\hline
\multirow{2}{*}{\textbf{Number of Peaks and Zero-Crossings}} 
    & Number of Max \& Min Peaks & Counts the local maxima and minima. A time point is a local maxima (minima) if both its two neighbouring time points are less (larger) than the current time point.\\
    & Number of Zero Crossings & Counts how often the signal crosses zero. \\
\hline
\end{tabular}
\caption{List of extracted features used for price time-series classification.\\
}
\label{tab:feature-summary}
\end{sidewaystable}

\begin{figure}
    \centering
    \includegraphics[width=1\linewidth]{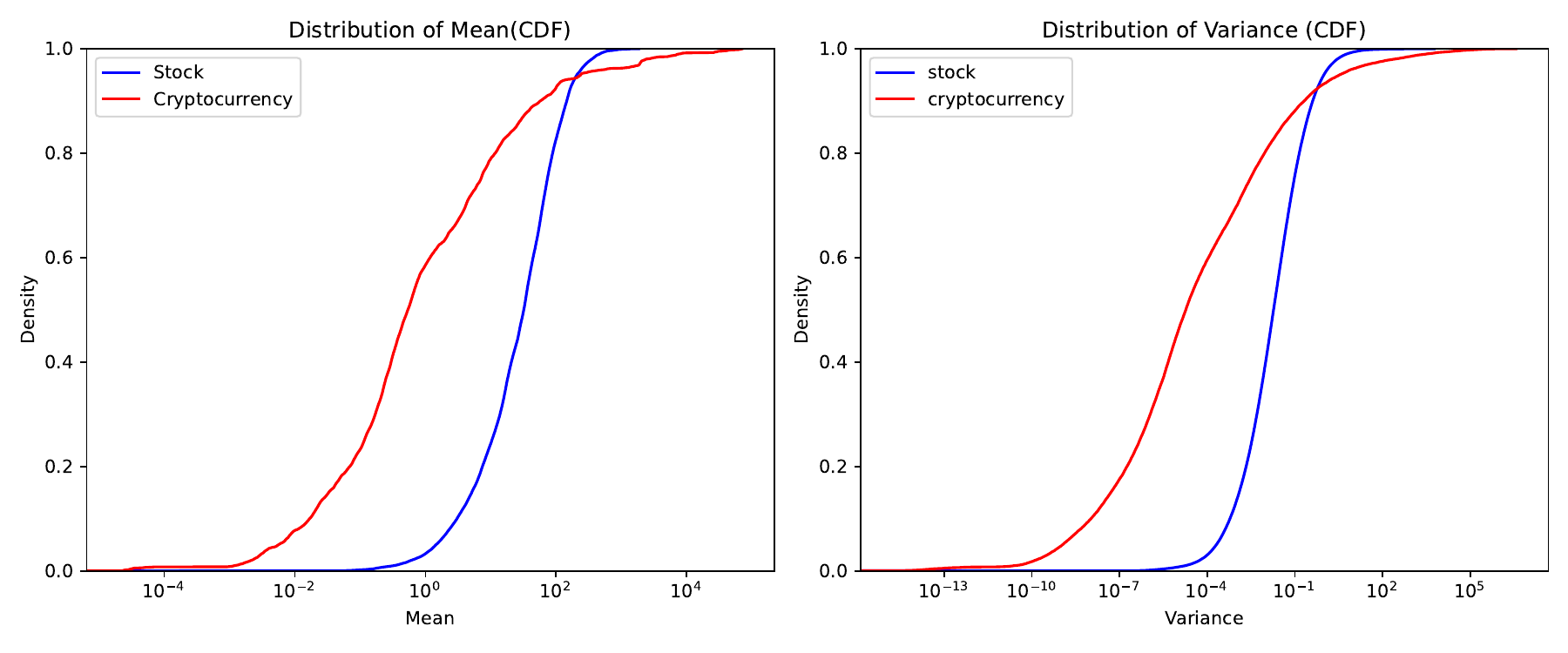}
    \caption{The CDF (cumulative distribution function) of mean and variance extracted from the original price time series data indicates that the mean and variance of cryptocurrencies fall in a broader range compared to stocks.}
    \label{fig:mean-var}
\end{figure}

\begin{figure}
    \centering
    \includegraphics[width=1\linewidth]{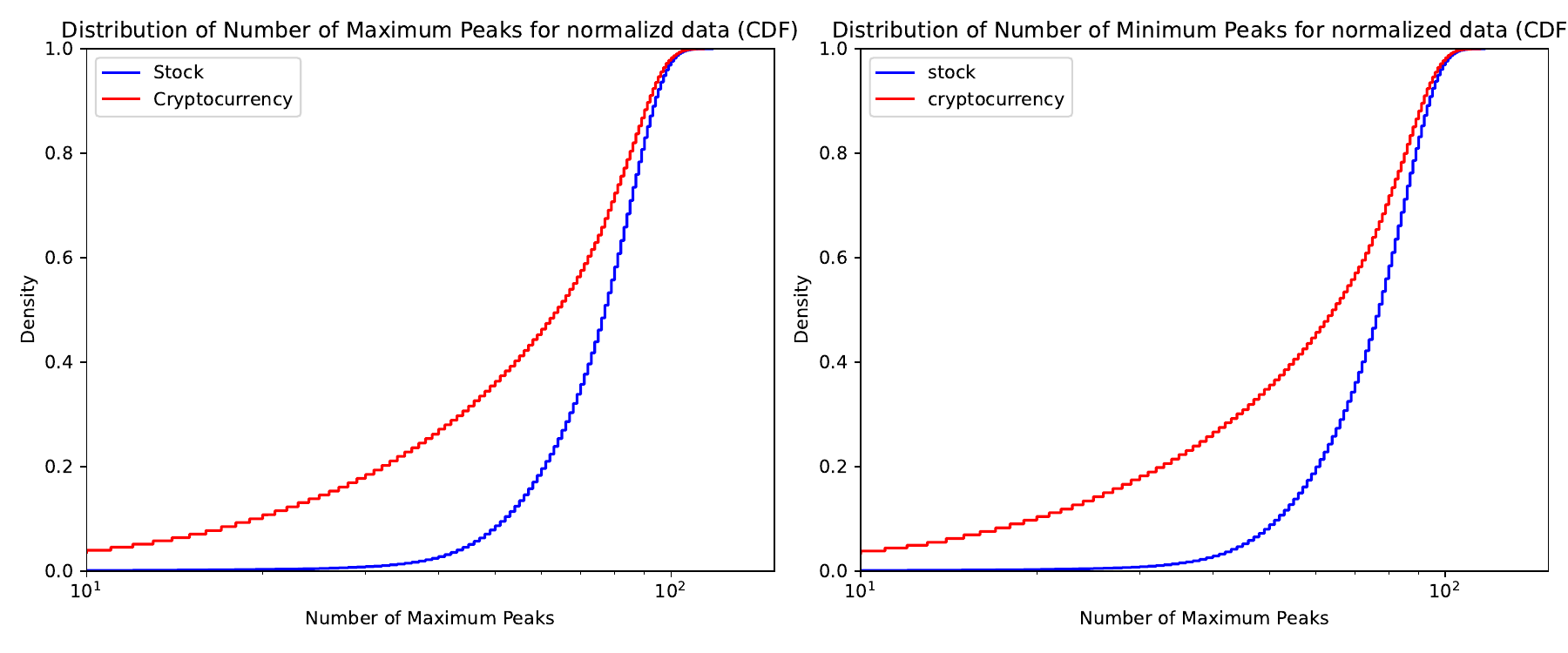}
    \caption{The CDF (cumulative distribution function) of the number of local maxima and minima (defined in Table \ref{tab:feature-summary}) extracted from normalized price data shows that stock data tends to exhibit more maximum or minimum peaks within a day compared to cryptocurrency data.}
    \label{fig:m-m-p-n}
\end{figure}

How important are these features in explaining the price time series differences? We take the mean, the variance, the count of maximum and minimum peaks of these two kinds of assets' price time series as examples for illustration. 
Figure \ref{fig:mean-var} shows the cumulative distribution function (CDF) of the mean and variance extracted from the original price time series data. Compared to stocks, the mean and variance of cryptocurrency fall in a broader range.
Features, such as the count of maximum and minimum peaks, show significant variations between cryptocurrencies and stocks, which is shown in Fig. \ref{fig:m-m-p-n}.

Now, we use machine learning methods to classify the cryptocurrency price time series and the stock price time series with these extracted features. The used machine learning methods in this experiment include logistic regression (LR), random forest (RF), support vector machine (SVM), k-nearest neighbors (KNN), and gradient boosting (GB), with corresponding classification results shown in table \ref{tab:feature1}.

The experiments in this section include 6 main sub-experiments. Firstly, we extracted corresponding features as shown in Table \ref{tab:feature-summary} from both the original cryptocurrencies' price time series and stocks' price time series, then the 5 machine learning methods (Logistic regression (LR), random forest (RF), support vector machine (SVM), k-nearest neighbors (KNN), and gradient boosting (GB)) were used on the extracted features to classify. Secondly, the 5 machine learning methods are used to classify on extracted features from the original return rate price time series (R), the normalized price time series (NP), and the normalized return rate time series (NR), respectively.
At last, for each price time series, we combine its features and the features of the corresponding return rate time series (P+R), and then we use the 5 machine learning methods to classify. The same experiment setting above also applies to the normalized price time series and normalized return rate time series (NP+NR). We have 6 sub-experiments in total, whose results are shown in Table \ref{tab:feature1}.

\begin{table}[h]
\centering
\renewcommand{\arraystretch}{1.5}
\footnotesize 
\begin{tabular}{ccccccc}
\hline
\textbf{Features extracted from} & \textbf{LR} & \textbf{RF}  & \textbf{SVM} & \textbf{KNN} & \textbf{GB} \\ \hline
Price Time Series (P)& 65.81\% & 94.37\%  & 66.86\% & 84.59\% & 91.63\%\\ 
Return Rate Time Series (R) & 70.02\% & 86.04\%  & 61.44\%  & 72.67\% & 84.58\%\\ 
Normalized Price Time Series (NP)& 65.68\% & 78.40\%  & 65.33\% & 63.92\% & 74.20\%\\ 
Normalized Return Rate Time Series (NR) & 69.46\% & 80.96\%  & 68.09\% & 70.86\% & 79.16\%\\ 
P+R & 64.81\% & 96.05\%  & 66.36\% & 87.92\% & 93.47\%\\
NP+NR  & 70.12\% & 84.88\%  & 71.21\% & 71.42\% & 82.88\% \\ 
\hline
\end{tabular}
\caption{Accuracy rate using features extracted from original price time series (P), the return rate time series (R), the normalized price time series (NP), and the normalized return rate time series (NR) with 5 machine learning models. 
P+R means that the experiment is conducted on feature data that are from both the original price time series and the original return rate time series. NP+NR denotes similarly to P+R.
Extracted features are those listed in Table \ref{tab:feature-summary}, which include time series' mean, variance, max, min, kurtosis, skewness, first, second, third-order auto-correlation, mean difference, mean absolute difference, peak-to-peak distance, entropy, area under the curve, number of maximum and minimum peaks and zero crossing rate. In these experiments, Logistic Regression (LR), Random Forest (RF), Support Vector Machine (SVM), K-Nearest Neighbors (KNN), and Gradient Boosting (GB) are employed in classification. }
\label{tab:feature1}
\end{table}

Among these methods, random forest (RF) consistently attained the best accuracy in both experiment settings. Gradient boosting (GB) was ranked as the second most effective method compared to other machine learning methods. For example, the accuracy rate with random forest (RF) and Gradient boosting (GB) is as high as 96\% and 93\%, respectively, in the P+R experiment setting, which means that the extracted features can help to explain much of the price pattern difference between cryptocurrencies and stocks. We also notice that the accuracy rate achieved by random forest (RF) and gradient boosting (GB) methods was close to that of neural network classification on the time series data when using features extracted from the original price time series data, which indicate the superior performance of the two machine learning models in classification based on several extracted features.

However, models such as SVM, LR, and KNN exhibited lower accuracy rates (64.8\% to 87.9\%), suggesting that the extracted features are insufficient to fully capture the differences in price patterns between cryptocurrencies and stocks. This implies the existence of additional, unaccounted-for features that may be critical.

Meanwhile, combining features extracted from the price time series and return rate time series could slightly improve the performance of these models, and the models performed better on features extracted from the original data than those from the normalized data.

\section{Conclusion and Discussion}\label{sec6}

In this paper, we applied various machine learning models to classify the price data of stocks and cryptocurrencies under different configurations and achieved remarkable classification results. All models' accuracy rates with different configurations achieve $>92\%$ in most data feed settings (``balanced" versus ``unbalanced", ``weekly" versus ``daily", ``four features" versus ``one feature"), except one setting with daily unbalanced data. 
The closed price time series can provide most information about the price time series' pattern difference of different assets when compared to all four price features (open price, the highest price, the lowest price, and closed price). The longer the price time series is, the more price pattern differences can be reflected, which can be seen from the two sub-experiments in Subsection \ref{4.1}, sub-experiment with daily unbalanced data and weekly unbalanced data.

When data used in training is balanced, the classification results improve by comparing the accuracy rate in Subsection \ref{4.2} and \ref{4.1}. The classification results with balanced data are also very robust, even if we change the machine learning models' architecture a lot.
The high accuracy rate and robust results suggest that the price time series data for stocks and cryptocurrencies may exhibit distinct patterns that reflect investors' different behaviors while holding and trading the two kinds of assets. However, the extremely high classification accuracy rate and F1 score may also originate from the superior performance of our used machine (deep) learning methods. To prove that the high accuracy is due to the price pattern difference between cryptocurrencies' price time series and stocks' price time series, but not due to the superior performance of the machine learning models, we use the same machine learning models to classify cryptocurrency price time series which are randomly labeled as 0 and 1 and get almost a 50\% test accuracy rate. The same experiment is repeated on the stock price time series, and we obtain an almost 50\% test accuracy rate. These two experiments help to prove that the extremely high accuracy rate and F1 score in Subsections \ref{4.1} and \ref{4.2} originate from the price pattern difference between cryptocurrencies and stocks.

Having established that cryptocurrency price time series exhibit distinct patterns compared to those of stocks, we extract features—such as mean, variance, maximum, minimum, kurtosis, skewness, and first to third-order autocorrelations—as summarized in Table \ref{tab:feature-summary}. We then apply machine learning methods including logistic regression (LR), random forest (RF), support vector machine (SVM), k-nearest neighbors (KNN), and gradient boosting (GB) for classification. The results indicate that these features partially explain the pattern differences: while sufficient for some models, they fall short for others, suggesting that additional critical features remain to be identified. Addressing this limitation will be the focus of future work.

\end{document}